\documentclass[a4paper,twoside,10pt]{article}

\input{jgrg19.sty}
\begin{document}
%
\pagestyle{fancy}
\fancyhead{}
  \fancyhead[RO,LE]{\thepage}
  \fancyhead[LO]{N. Kan}                  
  \fancyhead[RE]{BSs under DD}    
\rfoot{}
\cfoot{}
\lfoot{}
\label{P26}    
\title{%
  Boson Stars under Deconstruction
}
%
\author{%
  Nahomi Kan\footnote{Email address: kan@yamaguchi-jc.ac.jp}$^{(a)}$
  and
  Kiyoshi Shiraishi\footnote{Email address: shiraish@yamaguchi-u.ac.jp}$^{(b)}$
}
%
\address{%
  $^{(a)}$Yamaguchi Junior College,
   Hofu-shi, Yamaguchi 747-1232, Japan\\
  $^{(b)}$Yamaguchi University,
  Yamaguchi-shi, Yamaguchi 753-8512, Japan
  }
%
\abstract{
We study solutions for boson stars in multiscalar theory.
We start with simple models with $N$ scalar theories.
Our purpose is to study the models
in which the mass matrix of scalars and the scalar couplings are given 
by an extended method of dimensional deconstruction.
The properties of the boson stars are investigated 
by the Newtonian approximation with the
large coupling limit. 
}

\section{Introduction}
Boson Stars (BSs) have been studied
in expectation of  solving 
the rotation curve (RC) problem. 
Many authors have attempted to explain RC of  galaxies
by assuming the existence of the galactic scale BS.
In order to fit the observable data,
the mass density of BS needs to be widely distributed.
This configuration can be constructed by the models such as 
BSs with scalar particles in excited states or the rotating BSs, 
but these BSs are unstable.
Whereas
Newtonian BSs with all the particles in only one ground state is stable,
it is difficult to illustrate the realistic RC.
Alternative models to solve these problems have been studied by Matos and
Ure\~na-L\'opez~\cite{P26_MU07},  and recently by Bernal {\it
et~al.}~\cite{P26_BBAP09}. They considered the multi-state BS, {\it i.~e.}
scalar fields both in ground and in excited states, with no (quadratic)
self-couplings.\\ In the present work,
We consider multi-kind scalar BS, not multi-states.
We suggest several models.
The first model contains two scalar particles with self- and mutual-couplings.
\footnote{Interacting boson stars and Q-balls have been 
studied by Brihaye {\it et~al.}~\cite{P26_Brihaye1,P26_Brihaye2,P26_Brihaye3}.}
The second model is build up under dimensional deconstruction (DD).
This model contains $N$ scalar particles interacting with oneself and with adjacent scalars.
DD has an aspect of latticized extra dimensions,
and 
the latter model could be an alternative to a higher dimensional BS.
In each model, we examine BS with large coupling limit.
We also consider BSs under extended DD. 
In this model, DD is generalized to field theory based on a {\it graph},
and the interactions between scalar particles are restricted by supersymmetry (SUSY). 

\section{Non-Relativistic Multi-scalar Boson Star}
\begin{figure}[h]
\begin{center}
\begin{minipage}{68.5mm}
\begin{center}
\includegraphics[keepaspectratio=true,width=6cm]{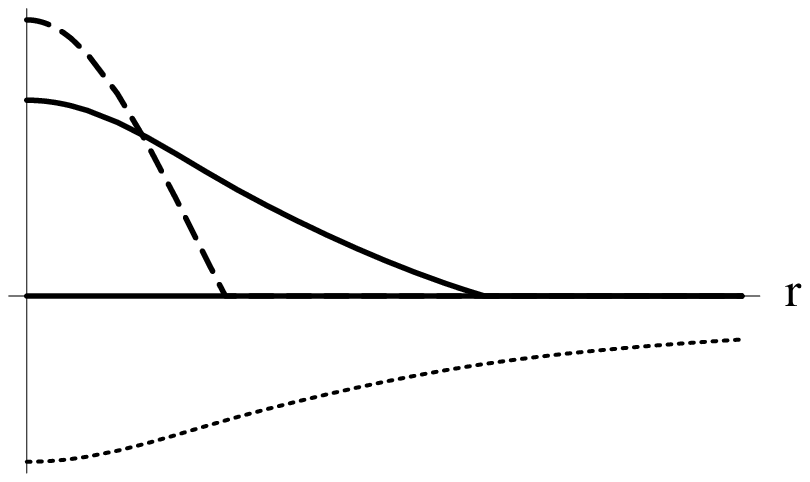}
\caption{
The behavior of the scalar fields $\Psi_1, \Psi_2$
and the gravitational potential $\phi$ as the function of the rescaled distance $r$.
The solid line, the broken line and the dotted line
represent $\Psi_1$, $\Psi_2$ and $\phi$, respectively.
The potential is spread out  by the existence of $\Psi_1$. 
}
\label{fig:P26_fig2_1}
\end{center}
\end{minipage}
\hspace{20mm}
\begin{minipage}{68.5mm}
\begin{center}
\includegraphics[keepaspectratio=true,width=6cm]{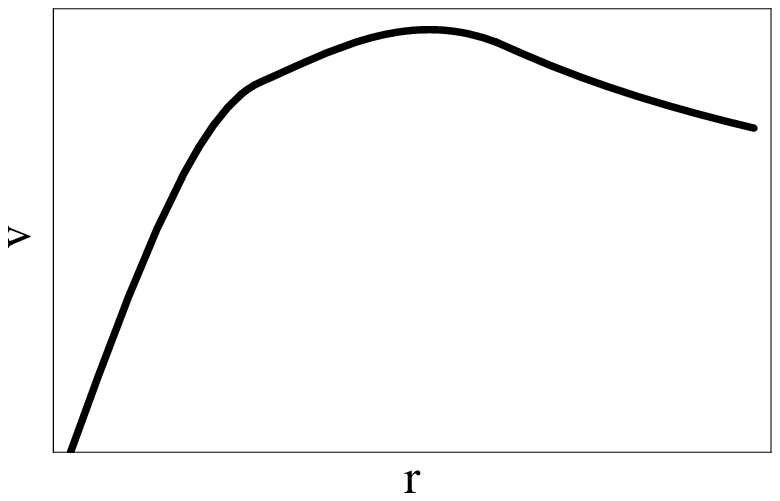}
\caption{
The behavior of the rotational velocity $v$ as the function of the rescaled distance $r$. 
The multi-scalar configuration improve RC.}
\label{fig:P26_fig2_2}
\end{center}
\end{minipage}
\end{center}
\end{figure}

\begin{figure}[h]
\begin{center}
\begin{minipage}{68.5mm}
\begin{center}
\includegraphics[keepaspectratio=true,width=6cm]{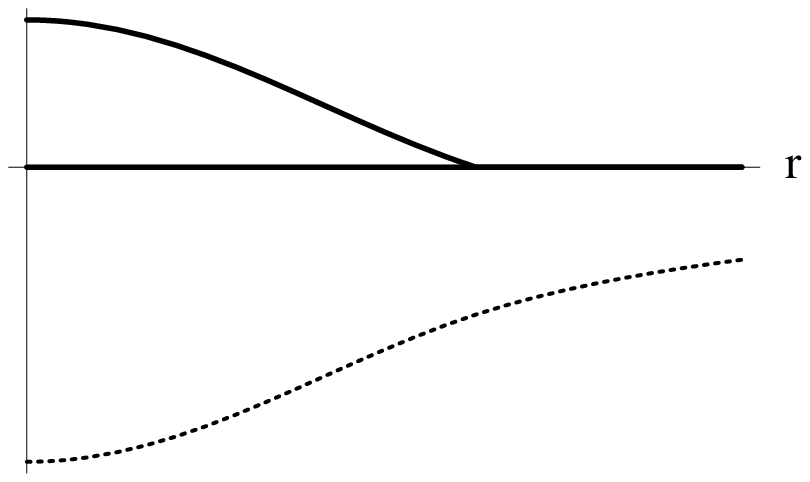}
\caption{
The behavior of the scalar field and the gravitational potential as the function of the rescaled distance $r$.
The solid line represents the scalar field,
whereas the dotted line represents the gravitational potential.
Compared  to the two scalar field model (Fig.~\ref{fig:P26_fig2_1}),
the range of the potential becomes narrow.
}
\label{fig:P26_fig1_1}
\end{center}
\end{minipage}
\hspace{20mm}
\begin{minipage}{68.5mm}
\begin{center}
\includegraphics[keepaspectratio=true,width=6cm]{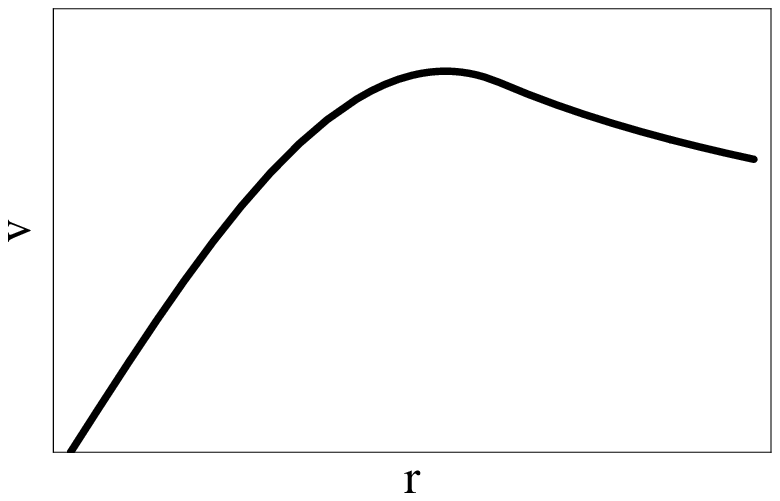}
\caption{
The behavior of the rotational velocity $v$ as the function of the rescaled distance $r$. 
The RC looks far from a satisfactory explanation of the observational data.
}
\label{fig:P26_fig1_2}
\end{center}
\end{minipage}
\end{center}
\end{figure}

We consider a BS model, in which two scalar particles $\psi_1$ and $\psi_2$
with self- and mutual-couplings  $g_{ij}$,
described by the Hamiltonian:
\begin{eqnarray}
\label{P26_Hamiltonian2}
H-\mu_1 \tilde{N_1}-\mu_2 \tilde{N_2} &=&
\frac{\hbar^2}{2m_1}|\nabla \psi_1|^2+\frac{\hbar^2}{2m_2}|\nabla \psi_2|^2
+ \left( m_1|\psi_1|^2+|\psi_2|^2  \right)\phi 
+  \frac{1}{8\pi G} (\nabla \phi)^2 \nonumber \\
&&- \mu_1|\psi_1|^2 - \mu_2|\psi_2|^2
+\frac{1}{4}\frac{\hbar^3}{c} 
\left( \frac{g_{11}}{{m_1}^2}|{\psi_1}|^4+2\frac{g_{12}}{m_1 m_2}|\psi_1|^2|\psi_2|^2 
+\frac{g_{22}}{{m_2}^2} |\psi_2|^4 \right),
\end{eqnarray}
where
$\tilde{N_i}$ and $\mu_i$ are the number density and the chemical potential of the $i$-th scalar,
respectively,
whereas 
 $\phi$ is the gravitational potential.
We also normalize the particle number to
$N_i =\int d^3r |\psi_i|^2$.
In the large coupling limit~\cite{P26_CSW86},
the equations of motion are as follows:
\begin{eqnarray}
&&~~~
\nabla^2 \phi = 4\pi G(m_1|\psi_1|^2+m_2|\psi_2|^2),\\
&&m_1 \phi \psi_1+ \frac{1}{2} \left( \frac{g_{11}}{{m_1}^2}|\psi_1|^2+\frac{g_{12}}{m_1 m_2}|\psi_2|^2\right)=\mu_1\psi_1,\\
&&m_2 \phi \psi_2 + \frac{1}{2} \left( \frac{g_{22}}{{m_2}^2}|\psi_2|^2+\frac{g_{12}}{m_1 m_2}|\psi_1|^2\right)=\mu_2\psi_2,
\end{eqnarray}
where $\hbar=c=1$.
In the core of the BS, where $\psi_1 \neq 0, \psi_2 \neq 0$,
 the gravitational potential becomes 
\begin{equation}
\phi +  const. \propto -\frac{\sin (\omega r)}{r}, 
\end{equation}
where $r$ is the distance from the center of the BS, and
\begin{equation}
\omega^2 = 8\pi G \frac{{m_2}^4 g_{11} - 2{m_1}^2 {m_2}^2 g_{12}+{m_1}^4 g_{22}}{g_{11} g_{22}-{g_{22}}^2}.
\end{equation}
The outside of the BS, where $\psi_1 \neq 0, \psi_2 = 0$,
 the gravitational potential is
\begin{equation}
\phi +  const. \propto -\frac{\sin (\omega_1 r + \delta)}{r},
\end{equation}
with
$\omega_1 = 8\pi G {m_1}^4/{g_{11}}^2$.
The typical structure of BS are shown in Fig.~\ref{fig:P26_fig2_1} 
and the rotational curves in Fig.~\ref{fig:P26_fig2_2}.
We can find that the gravitational potential is spread out  by the existence of $\Psi_1$,
and which leads to an improvement of RC.
%
%
%
%
If a single scalar field model is considered,
which realized by $\Psi_2=0$ in (\ref{P26_Hamiltonian2}),
the range of the gravitational potential becomes narrow (Fig.~\ref{fig:P26_fig1_1}),
and 
the RC looks far from a satisfactory explanation of the observational data (Fig.~\ref{fig:P26_fig1_2}).
%
%
%
%
\section{General-Relativistic Boson Star under Deconstruction}
We propose three models of BSs under the DD scheme.
\subsection{Static and Spherical Boson Star}
We consider self-interacting $U(1)$ scalar field theory in DD.
This model is described by the action:
\begin{equation}
S_B = \int d^4 x \sqrt{-g} \sum_{i=1}^{N}
\left\{ |\partial_\mu \phi_i|^2 - m^2|\phi_i|^2 -f^2|\phi_{i+1}-\phi_{i}|^2 - \frac{ \tilde{\lambda} N}{2}|\phi_i|^4\right\}.
\end{equation}
If $f=0$, 
$[U(1)]^N$ symmetry recovers.
Ansatz  for a static and spherical BS: 
$\phi_i (x) = \frac{1}{\sqrt{N}} \phi (r) e ^{-i \omega t +i \theta_i}$,
where $\theta_i$ is a constant number,
leads to
the square of a scalar boson mass:
\begin{equation}
m^2+\frac{f^2}{N} \sum^N_{i=1}|1-{\rm e}^{i(\theta_{i+1}-\theta_{i})}|^2 \equiv m_b^2 ,
\end{equation} 
and to 
the charge density:
\begin{equation}
\sum^N_{i=1}i(\phi_{i}^{\ast}\partial_{0} \phi_{i}- \phi_{i}\partial_{0}\phi_{i}^{\ast})=2\omega\phi(r)^2,
\end{equation} 
where the charge are equally distributed on each site.
The BS mass made of a single scalar with mass $m_b$ becomes
 $M_{BS} \sim \frac{M_{pl}^2}{m_{b}}$
for no coupling, 
whereas  
$M_{BS} \sim \sqrt{\tilde{\lambda}}\frac{M_{pl}^{3}}{m_{b}^{2}}$
for large coupling,
where we use the convention of Jetzer~\cite{P26_J92}.
Stars with these maximum masses are stable.
If $\theta_{1}=\cdots=\theta_{N}$, or $N\to \infty$,
Kaluza-Klein (KK) theory is recovered
and
BSs are made of the zero-mode field
with a minimum mass $(m_{b}^{2})_{min}=m^2$.
 \subsection{Boson Star under Generalized Deconstruction --SUSY-inspired model--}
We extend DD to the model based on graph theory. 
In this model,
a continuum limit is not necessary,
and 
$U(1)$ interactions at each site (vertex) of an arbitrary graph is still invariant.
We also assume SUSY in order to restrict the other interactions
\cite{P26_KKS09},
and then simplify the interaction terms.
This model is described by the action:
\begin{equation}
S_B = \int d^4 x \sqrt{-g} \sum_{i, j=1}^{N}
\left\{ |\partial_\mu \phi_i|^2 - m^2|\phi_i|^2 -f^2\phi_{i}^{\ast}\triangle_{ij}\phi_{j} 
- \frac{N\Lambda_{ij}}{2}|\phi_i|^2|\phi_j|^2\right\} ,
\end{equation}
where $\lambda = \sum_{i, j}\Lambda_{i j}$, and $\triangle$ is graph laplacian.
Unfortunately, this model also describes a single $U(1)$ charge in general.
Thus the most probable BS in this model is made of a scalar field with the minimum mass.

 \subsection{Graph-oriented model}
Characteristic matrices associated with a graph are the graph laplacian $\triangle$
and identity matrix.
We expand the scalar field by the eigenvector of the graph laplacian, 
such as
$\vec{\phi}=\{ \phi_{1}, \phi_{2}, \cdots , \phi_{N} \} = \sum_{a=1}^{N} \phi_{a}\vec{x}_{a}$,
where
$\triangle\vec{x}_{a}=\lambda_{a}\vec{x}_{a}$, 
and
$\vec{x}_{a} \cdot \vec{x}_{b}=\delta_{ab}$,
as usual.
In this notations,
BS based on the graph with $p$ vertices and $q$ edges
is described by the action:
\begin{equation}
S_B = \int d^4 x \sqrt{-g}
\left\{
\partial \vec{\phi}^{\dag} \cdot \partial \vec{\phi} - m^2 |\vec{\phi}|^2 -f^2\vec{\phi}^{\dag} \triangle \vec{\phi} 
-\frac{\Lambda_{p}}{2}(|\vec{\phi}|^2)^2
- \Lambda_q |\vec{\phi}|^2 \vec{\phi}^{\dag} \triangle \vec{\phi} 
- \frac{\Lambda_{r}}{2}(\vec{\phi}^{\dag} \triangle\vec{\phi})^2
\right\}.
\end{equation}
If $\phi_{3}=\cdots=\phi_{N}=0$ and $\lambda_{1}=0$,
a boson mass becomes $m_{a} = \sqrt{m^2+f^2 \lambda_{a}}$
and interaction terms are
\begin{equation}
\label{P26_GOM_int2}
-\frac{\Lambda_p}{8}\left( \frac{|\psi_1|^2}{m_1}+\frac{|\psi_2|^2}{m_2} \right)^2
-\frac{\Lambda_q}{4}\lambda_2 \frac{|\psi_2|^2}{m_2}
\left( \frac{|\psi_1|^2}{m_1}+\frac{|\psi_2|^2}{m_2} \right)
-\frac{\Lambda_r}{8}{\lambda_2}^2 \frac{|\psi_2|^4}{m_2} ,
\end{equation}
where $\psi_a \equiv \sqrt{2m_a}\phi_a e^{i m_a t}$. 
Self- and mutual-couplings  $g_{ij}$
can be read
from (\ref{P26_GOM_int2}),
such as $g_{11}=\Lambda_q/4$, and so on.
By similar ways, we can obtain systematic construction of many-scalar models with several conserved charges.
\section{Summary and Outlook}
We have examined the Newtonian boson star with two $U(1)$ charges in the large-coupling limit.
The understanding of 
rotation curves of galaxies are improved in this model.
We have also examined BSs under deconstruction.
In the continuum limit, 
it is found that the possible BS is made of ``zero mode'' field. 
We have suggested two models of BSs based on graph theory.\\
As future work,
we will consider
general relativistic BSs and graph-oriented models 
with many charges or arbitrary couplings.
We will also investigate excited states
under the condition of fixing the mass and size of BSs.
Time dependent solutions or oscillations are also interesting.
We wish to study these subjects, elsewhere.  
\section*{Acknowledgements}
The authors would like to thank K. Kobayashi for useful comments, and also the organizers of JGRG19.

\end{document}